\documentclass[a4paper, UKenglish, cleveref, autoref, thm-restate]{lipics-v2021}
\nolinenumbers

\usepackage{cmll}
\usepackage{xcolor}
\usepackage{framed}

\usepackage{quiver}

\newcommand\inputs[1]{{\color{red}{#1}}}
\newcommand\outputs[1]{{\color{blue}{#1}}}
\newcommand\mumut{\mu\tilde{\mu}}
\newcommand\chk{\mathrm{chk}}
\newcommand\syn{\mathrm{syn}}
\newcommand\term{\mathrm{tm}}
\newcommand\coterm{\mathrm{cotm}}
\newcommand\cmd{\mathrm{cmd}}
\newcommand\match{\mathrm{match}}
\newcommand\comatch{\mathrm{comatch}}

\title{Canonical bidirectional typechecking}

\author{Zanzi Mihejevs}{Glasgow Lab for AI Verification}{}{}{}
\author{Jules Hedges}{Glasgow Lab for AI Verification}{}{}{}
\authorrunning{Z. Mihejevs and J. Hedges}
\Copyright{Zanzi Mihejevs and Jules Hedges}

\begin{document}

\maketitle

\begin{abstract}
	We demonstrate that the checkable/synthesisable split in bidirectional typechecking coincides with existing dualities in polarised System L, also known as polarised $\mumut$-calculus. Specifically, positive terms and negative coterms are checkable, and negative terms and positive coterms are synthesisable. This combines a standard formulation of bidirectional typechecking with Zeilberger's `cocontextual' variant. We extend this to ordinary `cartesian' System L using Mc Bride's co-de Bruijn formulation of scopes, and show that both can be combined in a linear-nonlinear style, where linear types are positive and cartesian types are negative. This yields a remarkable 3-way coincidence between the shifts of polarised System L, LNL calculi, and bidirectional calculi.
\end{abstract}

\section{Introduction}

Bidirectional typechecking is a methodology for designing typecheckers by splitting syntax into fragments whose types are \emph{checkable} and \emph{synthesisable}, also called \emph{inferrable} \cite{dunfield2021bidirectional}. While well-established `recipes' for developing bidirectional typing disciplines exist \cite{dunfield_pfenning_tridirectional,dunfield2021bidirectional}, extending these in new directions is still more art than science. In this paper we demonstrate that a \emph{canonical} bidirectional typing discipline is possible - that is, no choices need to be made - for the polarised classical type theory System L, also known as $\mu\tilde\mu$-calculus. It has long been observed that there is a connection between polarisation and bidirectional typechecking \cite{semanticdomain,conor,mercer2022implicit}, but the precise relationship has remained unclear. 

Polarised System L \cite{curien2000duality,munch2013syntax,downen2017sequent,ostermann2022introduction,binder2024grokking} is a type theory that combines both polarity and \emph{chirality} - the positive fragment is driven by cuts between a term and a pattern, and the negative fragment is driven by cuts between a \emph{coterm} and a \emph{copattern}.

Remarkably, linear System L \cite{munch2009focalisation, spiwack2014dissection} admits a canonical bidirectional typing discipline based on a combination of ideas from both standard and co-contextual typing \cite{zeilberger2015balanced,erdweg2015co,krishnaswami-inverting}, giving us a ``bi-contextual'' typing algorithm. This lets us equip a type system based on polarised linear logic with a bidirectional discipline where all typing annotations are exclusively limited to shifts between synthesisable and checkable judgements. Although linearity plays a crucial role, we are able to drop the linearity restriction by using a very specific scoping discipline known as \emph{co-de Bruijn} \cite{mcbride-everybodys-somewhere}.

We demonstrate that polarised System L requires almost no modifications to make it fit into this bi-contextual discipline. The checkable/synthesisable mode of each connective is fully determined by the sort of its judgement in polarised System L, and the only modification that we need to do is to add annotations on two of the rules for System L's shift modalities. This coincides with where we would be required to put typing annotations on one of the bidirectional shifts in a standard presentation of bidirectional $\lambda$-calculus. That is to say, the \emph{logical} shifts of System L essentially coincide with the \emph{bidirectional} shifts of $\lambda$-calculus.

Overall there are three axes of symmetry: \emph{polarity} (positive and negative), \emph{chirality} (term and coterm, also called producer and consumer) and \emph{mode} (checkable and synthesisable), and \emph{any two determine the third}. In practice the first two are canonical, so we use polarity and chirality to determine the typechecking mode. The following table illustrates this:

\begin{tabular}{|c|c|c|}
	\hline
	& Term & Coterm \\
	\hline
	Positive & Checkable expression & Synthesisable pattern \\
	\hline
	Negative & Synthesisable copattern & Checkable coexpression \\
	\hline
\end{tabular}


To typecheck the positive fragment, we start with a cut between an expression (positive term) and a pattern (positive coterm).
\[ \frac{\Gamma\vdash \text{producer} \Leftarrow A\ |\ \Delta \qquad \Gamma'\ |\ \text{pattern} \Rightarrow A\vdash \Delta'}{\left<\text{producer } | \text{ pattern}\right> : (\Gamma, \Gamma' \vdash \Delta, \Delta') } (\text{cut}^+) \]
We first synthesise the type of the pattern, then check the type of the expression against the synthesised type, after which we are done.

Dually, to typecheck the negative fragment, we start with a cut between a coexpression (negative coterm) and a corresponding copattern (negative term).
\[ \frac{\Gamma\ |\ \text{copattern} \Rightarrow A\vdash \Delta \qquad \Gamma'\vdash \text{consumer} \Leftarrow A\ |\ \Delta'}{\left<\text{copattern } | \text{ consumer}\right> : (\Gamma, \Gamma' \vdash \Delta, \Delta') } (\text{cut}^-) \]
We first synthesise the type of the copattern, then check the type of the coexpression against the synthesised type, after which we are done.

We draw attention to the fact that both the positive fragment and the negative fragment considered in isolation can be typechecked with no typing annotations whatsoever inside the syntax; annotations are only required when we combine the fragments together. Each of these fragments in isolation is not expressive enough to include function types and thus do not embed $\lambda$-calculus, although they do have nontrivial binders. The consequence of this is that typechecking every connective \emph{except} for functions is straightforward, which is the opposite of the usual situation for $\lambda$-calculi where typechecking functions is standard but extensions with nontrivial binders are difficult.

Extending to the full polarised calculus, we need to add negations and shifts. Curiously, negation inverts both the polarity and chirality of each judgement, but it does not change their bidirectional mode - so an expression becomes a coexpression and vice versa, and patterns swap with copatterns. This means that it does not require annotations.

The most interesting thing happens when we consider the shifts. The shifts amount to swapping \emph{both} the polarity and typechecking mode of a judgement. On one side, a copattern is shifted to an expression, and a pattern shifts to a coexpression. This corresponds to embedding a synthesisable judgement into a checkable one, and can be done with no issue (requiring a type equality or more generally a subtyping check). But the other way around, embedding a (co)expression into a (co)pattern, requires us to synthesise the type of a checkable judgement, which is precisely where we need to place a typing annotation.

Finally we extend System L to a linear-nonlinear calculus \cite{benton1994mixed} that we call System LNL, that can define the linear logic modalities $!$ and $?$ via adjoint modalities. We find that 2 of the 4 modalities coincide with the existing shifts of System L - which are also bidirectional shifts - while the other 2 are new.

The major finding of this paper is that canonical bidirectional typechecking is possible for a highly symmetric classical calculus such as polarised System L. However, a major motivation for us is to develop a methodology to design typecheckers for `ordinary' calculi such as dependent type theories. Our hypothesis is that designing a typechecking discipline for a language should amount to choosing an embedding into System L, System LNL or other similar systems, \emph{with no other choices needing to be made}. Such embeddings are generally well understood \cite{downen2018beyond}, and so we expect to obtain typing disciplines that could be named after the choice of embedding, such as a `call-by-value typesystem' or a `call-by-name typesystem'. However, we leave the details of this for future work.

\section{A standard bidirectional $\lambda$-calculus}

We begin with a standard presentation of bidirectional simply typed $\lambda$-calculus. 
The fundamental idea of bidirectional typing is that we separate term judgements into two classes, called \emph{checkable} and \emph{synthesisable} (or \emph{inferrable}). In a standard presentation:
\begin{itemize}
	\item Variables are synthesisable
	\item Abstractions are checkable if the body is checkable
	\item Applications are synthesisable if the function is synthesisable and the argument is checkable
\end{itemize}

The syntax of $\lambda$-calculus is extended with \emph{shifts} between the two classes. Any synthesisable term can be \emph{demoted} to merely checkable, and a checkable term can be \emph{promoted} to synthesisable at the expense of a type annotation inside the term.

We present this by two mutually defined judgements $\chk$ and $\syn$ that characterise the syntax of well scoped terms in a context of free variables. The scoping rules are given in figure \ref{fig:standard-lambda-scoping}. We draw attention to the standard practice that the contexts are shared between hypotheses, which we will be dropping in later sections.

\begin{figure}[h!]
	\begin{framed}
		\[ \frac{}{x_1, \ldots, x_n \vdash x_i\ \syn} \qquad\qquad \frac{\Gamma \vdash t\ \syn}{\Gamma \vdash\ \downarrow t\ \chk} \qquad\qquad \frac{\Gamma \vdash t\ \chk}{\Gamma \vdash\ \uparrow (t :: A)\ \syn} \]
		\[ \frac{\Gamma, x \vdash t\ \chk}{\Gamma \vdash \lambda x . t\ \chk} \qquad\qquad \frac{\Gamma \vdash t\ \syn \qquad \Gamma \vdash u\ \chk}{\Gamma \vdash tu\ \syn} \]
	\end{framed}
	\caption{Scoping rules of standard bidirectional $\lambda$-calculus}
	\label{fig:standard-lambda-scoping}
\end{figure}

The inputs and outputs of typechecking the two term sorts, and our notation for the corresponding typing judgements, are described in the following table:

\begin{tabular}{|c|c|c|c|}
	\hline
	$\lambda$-calculus & Typechecker inputs & Typechecker outputs & Notation \\
	\hline
	Checkable terms & typed context, type, checkable term & - & $\inputs\Gamma \Rightarrow \inputs A \ni \inputs t$ \\
	\hline
	Synthesisable terms & typed context, synthesisable term & type & $\inputs\Gamma \Rightarrow \inputs t \in \outputs A$ \\
	\hline
\end{tabular}

Following Mc Bride \cite{mcbride-ni}, we use the symbols $\in$ and $\ni$ (pronounced ``in'' and ``ni'' respectively) to denote type synthesis and checking, to maintain the lexical principle that the information flow in the typechecker is written left-to-right in judgements. We also replace the turnstile with an implication arrow, for reasons that will become increasingly clear over the course of the next several sections.

Throughout this paper we will consistently use the colours red and blue to describe inputs and outputs of a typechecker respectively. This is always from an `outside' perspective rather from the perspective of a particular proof rule (we are keeping in mind Mc Bride's principle that ``a proof rule is a client for its hypotheses and a server for its conclusion'' \cite{mcbride-iso}), and so from the perspective of a proof rule, the inputs that it \emph{receives} are red and the inputs that it \emph{provides} to typecheck a subterm are also red. Colour change from red to blue happens if we return one of our inputs, and colour change from blue to red happens if we take an output that we received from a hypothesis and then use it as an input to another hypothesis.

We will now describe the standard rules of bidirectional $\lambda$-calculus. Although they are presented as proof rules, they also implicitly describe an \emph{algorithm} in a syntax-directed way.

\begin{itemize}
	\item To synthesise a variable, we look it up in our input context:
	\[ \frac{}{\inputs{x_1 : A_1, \ldots, x_n : A_n} \Rightarrow \inputs{x_i} \in \outputs{A_i}} (\mathrm{Var}) \]

	\item To check that an abstraction $\lambda x . t$ has the type $A \to B$, we extend the context with $x : A$ and then check that $t$ has type $B$:
	\[ \frac{\inputs{\Gamma, x : A} \Rightarrow \inputs B \ni \inputs t}{\inputs \Gamma \Rightarrow \inputs{A \to B} \ni \inputs{\lambda x . t}} (\to\mathrm{-intro}) \]

	\item To synthesise the type of $tu$, we synthesise the type of $t$, deconstruct it as $A \to B$, check that $u$ has type $A$, and then return $B$:
	\[ \frac{\inputs\Gamma \Rightarrow \inputs t \in \outputs{A \to B} \qquad \inputs\Gamma \Rightarrow \inputs A \ni \inputs u}{\inputs\Gamma \Rightarrow \inputs{tu} \in \outputs B} (\to\mathrm{-elim}) \]

	\item To synthesise the type of an upshift, check the type annotation and then return it:
	\[ \frac{\inputs\Gamma \Rightarrow \inputs A \ni \inputs t}{\inputs\Gamma \Rightarrow \inputs{\uparrow\! (t :: A)} \in \outputs A} (\uparrow) \]

	\item To check that a downshift $\downarrow t$ has some type, we synthesise the type of $t$ and check that it agrees:
	\[ \frac{\inputs\Gamma \Rightarrow \inputs t \in \outputs B \qquad \inputs A \sqsubseteq \inputs B}{\inputs\Gamma \Rightarrow \inputs A \ni \inputs{\downarrow t}} (\downarrow) \]
\end{itemize}

The rule for downshifts makes use of an unspecified decidable relation $\sqsubseteq$ between types, which we informally think of as a subtyping relation. For simple type theories this can be equality of types, but there are several things that can be done here, including unification and genuine subtyping. We will return to this point in later sections.

We will now extend this basic $\lambda$-calculus with product and coproduct types. The scoping rules are given in figure \ref{fig:standard-lambda-scoping-2}. For the standard presentation, product introduction (pairing) is checkable and elimination (projections) is synthesisable, while coproduct introduction (injections) and elimination (case matching) are both checkable. It is possible to make nullary and binary product introductions synthesisable rather then checkable, but a standard presentation makes all introduction forms checkable and \emph{most} elimination forms synthesisable.

\begin{figure}[h!]
	\begin{framed}
		\[ \frac{}{\Gamma \vdash \left< \right>\ \chk} \qquad\qquad \frac{\Gamma \vdash t\ \chk}{\Gamma \vdash \mathrm{absurd}\ t\ \chk} \]
		\[ \frac{\Gamma \vdash t\ \chk \qquad \Gamma \vdash u\ \chk}{\Gamma \vdash \left< t, u \right>\ \chk} \qquad\qquad \frac{\Gamma \vdash t\ \syn}{\Gamma \vdash \pi_1 t\ \syn} \qquad\qquad \frac{\Gamma\vdash t\ \syn}{\Gamma \vdash \pi_2 t\ \syn} \]
		\[ \frac{\Gamma \vdash t\ \chk}{\Gamma \vdash \iota_1 t\ \chk} \qquad\qquad \frac{\Gamma \vdash t\ \chk}{\Gamma \vdash \iota_2 t\ \chk} \qquad\qquad \frac{\Gamma \vdash t\ \syn \qquad \Gamma, x \vdash u_1\ \chk \qquad \Gamma, y \vdash u_2\ \chk}{\Gamma \vdash \mathrm{case}\ t\ \mathrm{of}\ \{  \iota_L x \Rightarrow u_1; \iota_R y \Rightarrow u_2 \}\ \chk} \]
	\end{framed}
	\caption{Scoping rules for products and coproducts in standard bidirectional $\lambda$-calculus}
	\label{fig:standard-lambda-scoping-2}
\end{figure}

\begin{itemize}
	\item To check a nullary product introduction we only accept the unit type:
	\[ \frac{}{\inputs\Gamma \Rightarrow \inputs 1 \ni \inputs{\left< \right>}} (1\mathrm{-intro}) \]

	\item To check a binary product introduction we check both subterms:
	\[ \frac{\inputs\Gamma \Rightarrow \inputs A \ni \inputs t \qquad \inputs\Gamma \Rightarrow \inputs B \ni \inputs u}{\inputs \Gamma \Rightarrow \inputs{A \times B} \ni \inputs{\left< t, u \right>}} (\times\mathrm{-intro}) \]

	\item To synthesise a projection we synthesise whole type and return the appropriate part:
	\[ \frac{\inputs \Gamma \Rightarrow \inputs t \in \outputs{A \times B}}{\inputs\Gamma \Rightarrow \inputs{\pi_L t} \in \outputs A} (\times\mathrm{-elim-L}) \qquad\qquad \frac{\inputs \Gamma \Rightarrow \inputs t \in \outputs{A \times B}}{\inputs\Gamma \Rightarrow \inputs{\pi_R t} \in \outputs B} (\times\mathrm{-elim-R}) \]

	\item To check a nullary coproduct elimination we check that the subterm has the zero type:
	\[ \frac{\inputs\Gamma \Rightarrow \inputs 0 \ni \inputs t}{\inputs\Gamma \Rightarrow \inputs A \ni \inputs{\mathrm{absurd}\ t}} (0\mathrm{-elim}) \]

	\item To check an injection we check that the subterm has the appropriate type:
	\[ \frac{\inputs\Gamma \Rightarrow \inputs A \ni \inputs t}{\inputs \Gamma \Rightarrow \inputs{A + B} \ni \inputs{\iota_L t}} (+\mathrm{-intro-L}) \qquad\qquad \frac{\inputs\Gamma \Rightarrow \inputs B \ni \inputs t}{\inputs\Gamma \Rightarrow \inputs{A + B} \ni \inputs{\iota_R t}} (+\mathrm{-intro-R}) \]

	\item The binary coproduct elimination rule is quite complicated (and breaks a desired rule that elimination forms are always synthesisable): we synthesise the type of the scrutinee and then check the other subterms in the resulting extended contexts:
	\[ \frac{\inputs\Gamma \Rightarrow \inputs t \in \outputs{A + B} \qquad \inputs{\Gamma, x : A} \vdash \inputs C \ni \inputs{u_1} \qquad \inputs{\Gamma, y : B} \Rightarrow \inputs C \ni \inputs{u_2}}{\inputs\Gamma \Rightarrow \inputs C \ni \inputs{\mathrm{case}\ t\ \mathrm{of}\ \{ \iota_L x \Rightarrow u_1; \iota_R y \Rightarrow u_2 \}}} (+\mathrm{-elim}) \]
\end{itemize}

\section{Cocontextual linear $\lambda$-calculus}

Traditionally, bidirectional typechecking proceeds top-down, starting with a variable typing context, and so the context itself can be considered checkable. This works well because the main binding form in the $\lambda$-calculus is $\lambda$-abstraction, which is checkable, meaning that we can easily extend our context when going under a $\lambda$ binder.

However, things become less clear when working with binders inside of synthesisable terms. For instance, the positive definition of products is eliminated by a construct which binds two variables in the context. Since elimination forms are typically synthesisable, this means that we need to make the positive product elimination binder synthesisable. However, this means that we need some way to extend our typing context when going under the binder, and we can't do that because we haven't synthesised the required type yet.

Zeilberger \cite{zeilberger2015balanced} observed that these binding forms can be synthesised in what he called a dual bidirectional typing discipline, which is now known as cocontextual typechecking \cite{erdweg2015co}. Rather than starting with a type and a checkable variable context, it is enough to start with a type and a term, and then \emph{synthesise} the variable context during typechecking. In cocontextual typechecking this is possible because of two core assumptions: that the variable rule in a linear calculus has a singleton context, and that the rule is checkable rather than synthesisable. What this means is that by the time we reach a checkable variable leaf in the term, there are no other variables in the context, and so we trivially `check' that the variable matches the given type by synthesising its typing context. 


For the majority of cases, cocontextual typechecking swaps the checkable and synthesisable terms, so for example we consider variables and abstractions to be synthesisable and applications to be checkable. The exception is the shift rules, since we still need a type annotation to go from checkable to synthesisable but not the other way.

The well scoped linear $\lambda$-terms are generated by the rules given in figure \ref{fig:linear-lambda-scoping}. Notice that the scoping rules for variables and applications have changed to account for linearity. A variable is well scoped if the scope consists of exactly that variable, and the scope of an application is the disjoint union of the scopes of the two subterms.

\begin{figure}[h!]
	\begin{framed}
		\[ \frac{}{x \vdash x\ \chk} \qquad\qquad \frac{\Gamma \vdash t\ \syn}{\Gamma \vdash\ \downarrow t\ \chk} \qquad\qquad \frac{\Gamma \vdash t\ \chk}{\Gamma \vdash\ \uparrow (t :: A)\ \syn} \]
		\[ \frac{\Gamma, x \vdash t\ \syn}{\Gamma \vdash \lambda x . t\ \syn} \qquad\qquad \frac{\Gamma_1 \vdash t\ \chk \qquad \Gamma_2 \vdash u\ \syn}{\Gamma_1, \Gamma_2 \vdash tu\ \chk} \]
	\end{framed}
	\caption{Scoping rules for cocontextual linear $\lambda$-calculus}
	\label{fig:linear-lambda-scoping}
\end{figure}

In order to preserve our invariant that typechecker inputs go on the left and typechecker outputs go on the right, we are now required to write the context of a $\lambda$-term on the right hand side. To compensate, we replace the turnstile with a left arrow indicating the logical information flow. Thus checkable term judgements will be written $A \ni t \Leftarrow \Gamma$, and synthesisable term judgements will be written $t \in A \Leftarrow \Gamma$. In the following sections we will push this idea further, with contexts on both the left and right hand side.

Although it is possible to write a cocontextual typechecker that operates on ``raw'' terms, such a typechecker is required to also do a certain amount of \emph{scopechecking}. To avoid this, we choose to work with \emph{well scoped syntax} in which the linearity is intrinsic. 
Thus we require that the typechecker receives an additional input, which is the \emph{scoped} context containing the names (but not the types) of all variables.
We omit the scoped context inputs when writing judgements, but it is important to remember that they are there, and as a reminder we will write the \emph{names} of variables in synthesisable contexts in red (but their types in blue), locally breaking our principle that typechecker inputs appear to the left of outputs.

\begin{tabular}{|c|c|c|c|}
	\hline
	Cocontextual linear $\lambda$-calculus & Typechecker inputs & Typechecker outputs & Notation \\
	\hline
	Checkable terms & scoped context, type, checkable term & typed context & $\inputs A \ni \inputs t \Leftarrow \outputs\Gamma$ \\
	\hline
	Synthesisable terms & scoped context, synthesisable term & type, typed context & $\inputs t \in \outputs A \Leftarrow \outputs\Gamma$ \\
	\hline
\end{tabular}

The typing rules of cocontextual linear $\lambda$-calculus are given as follows:
\begin{itemize}
	\item To `check' that a variable $x$ has the type $A$, we synthesise the singleton context $x : A$:
	\[ \frac{}{\inputs A \ni \inputs x \Leftarrow \inputs x : \outputs A} (\mathrm{Var}) \]

	\item To synthesise the type and context of an abstraction $\lambda x . t$, we synthesise the type and context of $t$ and look up the type of $x$:
	\[ \frac{\inputs t \in \outputs B \Leftarrow \outputs\Gamma, \inputs x : \outputs A}{\inputs{\lambda x . t} \in \outputs{A \multimap B} \Leftarrow \outputs\Gamma} (\multimap\mathrm{-intro}) \]

	\item To check that an application has a given type, we synthesise the type of the argument and then check that the function has the appropriate function type:
	\[ \frac{\inputs u \in \outputs A \Leftarrow \outputs{\Gamma_2} \qquad \inputs{A \multimap B} \ni \inputs t \Leftarrow \outputs{\Gamma_1}}{\inputs B \ni \inputs{tu} \Leftarrow \outputs{\Gamma_1}, \outputs{\Gamma_2}} (\multimap\mathrm{-elim}) \]
	As written this rule is not syntax-directed; this is fixed using \emph{covers}, which we will introduce in the next section.

	\item The shift rules are essentially unchanged from before:
	\[ \frac{\inputs A \ni \inputs t \Leftarrow \outputs\Gamma}{\inputs{\uparrow (t :: A)} \in \outputs A \Leftarrow \outputs\Gamma} (\uparrow) \qquad\qquad \frac{\inputs t \in \outputs B \Leftarrow \outputs\Gamma \qquad \inputs A \sqsubseteq \inputs B}{\inputs A \ni \inputs{\downarrow t} \Leftarrow \outputs\Gamma} (\downarrow) \]
\end{itemize}

We can also add tensor (ie. multiplicative) products, for which both introduction and elimination is synthesisable. However, although linear $\lambda$-calculus can easily be extended with (additive) coproducts, giving them cocontextual typing rules requires subtyping. 


\begin{itemize}
	\item To synthesise a nullary product introduction, we return the unit type and empty context:
	\[ \frac{}{\inputs{\left< \right>} \in \outputs I \Leftarrow \cdot} (I\mathrm{-intro}) \]

	\item To synthesise a nullary product elimination, we check that the eliminated term has the unit type and synthesise the other subterm:
	\[ \frac{\inputs I \ni \inputs t \Leftarrow \outputs{\Gamma_1} \qquad \inputs u \in \outputs A \Leftarrow \outputs{\Gamma_2}}{\inputs{\mathrm{let} \left< \right> = t\ \mathrm{in}\ u} \in \outputs A \Leftarrow \outputs{\Gamma_1}, \outputs{\Gamma_2}} (I\mathrm{-elim}) \]

	\item To synthesise a linear product, we synthesise both subterms:
	\[ \frac{\inputs t \in \outputs A \Leftarrow \outputs{\Gamma_1} \qquad \inputs u \in \outputs B \Leftarrow \outputs{\Gamma_2}}{\inputs{\left< t, u \right>} \in \outputs{A \otimes B} \Leftarrow \outputs{\Gamma_1}, \outputs{\Gamma_2}} (\otimes\mathrm{-intro}) \]

	\item To synthesise a binary product elimination we first synthesise the right subterm, then we look up the bound variables in the synthesised typing context to check the left subterm:
	\[ \frac{\inputs u \in \outputs C \Leftarrow \outputs{\Gamma_2}, \inputs x : \outputs A, \inputs y : \outputs B \qquad \inputs{A \otimes B} \ni \inputs t \Leftarrow \outputs{\Gamma_1}}{\inputs{\mathrm{let} \left< x, y \right> = t\ \mathrm{in}\ u} \in \outputs C \Leftarrow \outputs{\Gamma_1}, \outputs{\Gamma_2}} (\otimes\mathrm{-elim}) \]
\end{itemize}

\section{Cocontextual co-de Bruijn $\lambda$-calculus}

\emph{Co-de Bruijn} is a scoping style developed by Mc Bride in \cite{mcbride-everybodys-somewhere} in which discarding of variables happens as early as possible, contrasting with de Bruijn scoping in which discarding happens as late as possible. Although originally presented in a nameless way (hence \emph{de Bruijn}), the essence of co-de Bruijn works equally well with names, as we use here. The central machinery of co-de Bruijn is two inductive relations on contexts, a ternary relation called \emph{covering} that is written $\Gamma_1 \uplus \Gamma_2 \sim \Gamma_3$ and handles copying, and a binary relation called \emph{thinning} that is written $\Gamma_1 \subseteq \Gamma_2$ and handles discarding. In this paper we need both relations for both scoped and typed contexts; we will give the rules only for typed contexts since the rules for scoped contexts are the same but with the types omitted.

The inductive rules for covering and thinning split into \emph{linear} and \emph{cartesian} rules. When restricted to just the linear rules, $\Gamma_1 \uplus \Gamma_2 \sim \Gamma_3$ witnesses that $\Gamma_3$ is an interleaving of $\Gamma_1$ and $\Gamma_2$, while $\Gamma_1 \subseteq \Gamma_2$ is the identity relation. Linear covering is precisely what we omitted from the previous section to simplify our presentation, that is required to make the typechecking rules syntax-directed.
\[ \frac{}{\cdot \uplus \cdot \sim \cdot} \qquad\qquad \frac{\Gamma_1 \uplus \Gamma_2 \sim \Gamma_3}{\Gamma_1, x : A \uplus \Gamma_2 \sim \Gamma_3, x : A} \qquad\qquad \frac{\Gamma_1 \uplus \Gamma_2 \sim \Gamma_3}{\Gamma_1 \uplus \Gamma_2, x : A \sim \Gamma_3} \]
\[ \frac{}{\cdot \subseteq \cdot} \qquad\qquad \frac{\Gamma_1 \subseteq \Gamma_2}{\Gamma_1, x : A \subseteq \Gamma_2, x : A} \]

To obtain cartesian covers and thinnings we add one additional rule to each, respectively for copying and discarding:
\[ \frac{\Gamma_1 \uplus \Gamma_2 \sim \Gamma_3}{\Gamma_1, x : A \uplus \Gamma_2, x : A \sim \Gamma_3, x : A} \qquad\qquad \frac{\Gamma_1 \subseteq \Gamma_2}{\Gamma_1 \subseteq \Gamma_2, x : A} \]

The general principles for how these relations are used in syntax are:
\begin{itemize}
	\item For every rule with two hypotheses, we include covers that describe how the contexts of the hypotheses combine into the contexts of the conclusion
	\item For every rule that binds a variable, we include a thinning that allows the bound variable to be immediately discarded if it will never be used
	\item Every other rule is written in a linear style
\end{itemize}

As a consequence of the third rule, we are forced to discard any unused variable immediately, because otherwise it will eventually reach a leaf rule, and since leaf rules are linear they must use all variables in their context. This is what allows us to extend cocontextual typechecking to a cartesian calculus without requiring subtyping.


The key difference between linear and co-de Bruijn $\lambda$-calculus are its scoping rules, which are given in figure \ref{fig:codebruijn-lambda-scoping}.

\begin{figure}
	\begin{framed}
		\[ \frac{}{x \vdash x\ \mathrm{chk}} \qquad\qquad \frac{\Gamma \vdash t\ \mathrm{syn}}{\Gamma \vdash\ \downarrow t\ \mathrm{chk}} \qquad\qquad \frac{\Gamma \vdash t\ \mathrm{chk}}{\Gamma \vdash\ \uparrow (t :: A)\ \mathrm{syn}} \]
		\[ \frac{\Gamma_2 \vdash t\ \mathrm{syn} \qquad \Gamma_2 \subseteq \Gamma_1, x}{\Gamma_1 \vdash \lambda x . t\ \mathrm{syn}} \qquad\qquad \frac{\Gamma_1 \vdash t\ \mathrm{chk} \qquad \Gamma_2 \vdash u\ \mathrm{syn} \qquad \Gamma_1 \uplus \Gamma_2 \sim \Gamma_3}{\Gamma_3 \vdash tu\ \mathrm{chk}} \]
	\end{framed}
	\caption{Co-de Bruijn $\lambda$-calculus, scoping rules}
	\label{fig:codebruijn-lambda-scoping}
\end{figure}

We will explain in detail how typechecking $\lambda$-abstraction and function application differs between linear and co-de Bruijn syntax. The typing rule for application is
\[ \frac{\inputs u \in \outputs A \Leftarrow \outputs{\Gamma_2} \qquad \inputs{A \to B} \ni \inputs t \Leftarrow \outputs{\Gamma_1} \qquad \inputs{\Gamma_1} \uplus \inputs{\Gamma_2} \sim \outputs{\Gamma_3}}{\inputs B \ni \inputs{tu} \Leftarrow \outputs{\Gamma_3}} (\to\mathrm{-elim}) \]
The difference between this rule and the linear function application rule (besides changing $\multimap$ to $\to$) is the syntax $\inputs{\Gamma_1} \uplus \inputs{\Gamma_2} \sim \outputs{\Gamma_3}$. When we are typechecking an application we already have access to its scoping information, which includes a cover $\Gamma_1 \uplus \Gamma_2 \sim \Gamma_3$ for the \emph{scoped} contexts, which we write in black. This is used twice during typechecking. Firstly it is used to split the scoped context $\Gamma_3$, which is a typechecker input, into the scoped contexts $\Gamma_2$ and $\Gamma_1$, which are inputs to the typechecker for the two hypotheses. Typechecking these hypotheses synthesises the associated \emph{typed} contexts $\outputs{\Gamma_2}$ and $\outputs{\Gamma_1}$. We then use the scoped cover again to combine these two typed contexts into a typed context $\outputs{\Gamma_3}$, which is our final output.

In the presence of the copy rule for cartesian covers, the operation of combining two typed contexts along a scoped cover can fail, because each of the two hypotheses could return incompatible types for a shared variable. The simplest possible thing is that we fail unless all types for the same variable are equal. In general we have a partial binary least upper bound operator $\wedge$ on types which we then extend to contexts. We believe that reifying this into a (total) intersection type former $\cap$ is an interesting direction for future work.

Overall, the definition of the operation to merge typed contexts along a scoped cover is given as follows:
\[ \frac{}{\inputs{\cdot} \uplus \inputs{\cdot} \sim \outputs{\cdot}} \qquad\qquad \frac{\inputs{\Gamma_1} \uplus \inputs{\Gamma_2} \sim \outputs{\Gamma_3}}{\inputs{\Gamma_1}, \inputs x : \inputs A \uplus \inputs{\Gamma_2} \sim \outputs{\Gamma_3}, \inputs x : \outputs A} \qquad\qquad \frac{\inputs{\Gamma_1} \uplus \inputs{\Gamma_2} \sim \outputs{\Gamma_3}}{\inputs{\Gamma_1} \uplus \inputs{\Gamma_2}, \inputs x : \inputs A \sim \outputs{\Gamma_3}, \inputs x : \outputs A} \]
\[ \frac{\inputs{\Gamma_1} \uplus \inputs{\Gamma_2} \sim \outputs{\Gamma_3}}{\inputs{\Gamma_1}, \inputs x : \inputs A \uplus \inputs{\Gamma_2}, \inputs x : \inputs B \sim \outputs{\Gamma_3}, \inputs x : \outputs{A \wedge B}} \]

Next, we give the typing rules for $\lambda$-abstraction. Unlike for linear $\lambda$-calculus this splits into a pair of rules, depending on the associated thinning which can either keep or discard the bound variable. If we keep the variable then the rule is
\[ \frac{\inputs t \in \outputs B \Leftarrow \outputs\Gamma, \inputs x : \outputs A \qquad \inputs \Gamma, \inputs x : \inputs A \subseteq \outputs \Gamma, \inputs x : \outputs A}{\inputs{\lambda x . t} \in \outputs{A \to B} \Leftarrow \outputs\Gamma} (\to\mathrm{-intro-keep}) \]
Just as for covers, we are using a \emph{scoped} thinning to operate on a \emph{typed} context. In general, we have a scoped thinning $\Gamma_1 \subseteq \Gamma_2$, although for $\lambda$-abstraction in particular we only require the case in which $\Gamma_2$ consists of $\Gamma_1$ extended with at most one variable. As a typechecker input we get a scoped context $\Gamma_2$, which we restrict along the thinning to a scoped $\Gamma_1$ by possibly discarding a variable. Then we typecheck the hypothesis which synthesises a typed $\outputs{\Gamma_2}$, and we finally \emph{extend} this along the thinning to a typed $\Gamma_1$. In the case that we did not discard a variable, both scoped restriction and typed extension are trivial.

In the case that we did discard the bound variable, we appear to be stuck: when we extend the typed context along the thinning we are required to synthesise a type for the unused variable from nothing. Thus even for a simply-typed calculus we are forced to use a limited form of subtyping. In particular we extend our type language with a top type $\top$ which is the type of an unused variable. We can now define the partial binary operator $\wedge$: it is the greatest lower bound for a discrete poset congruently extended with $\top$ as the maximal element $x \leq \top$. (By ``congruent'' we mean for example that if $A' \leq A$ and $B \leq B'$ then $A \to B \leq A' \to B'$.)

Equipped with the type $\top$ we can now give the other typing rule for $\lambda$-abstraction:
\[ \frac{\inputs t \in \outputs B \Leftarrow \outputs \Gamma \qquad \inputs \Gamma \subseteq \outputs \Gamma, \inputs x : \outputs \top}{\inputs{\lambda x . t} \in \outputs{\top \to B} \Leftarrow \outputs \Gamma} (\to\mathrm{-intro-drop}) \]

It is curious to notice that even in the presence of the $\top$ type, our binary operation $\wedge$ could still be an equality check. This is a subtle consequence of the invariants of co-de Bruijn scoping. When we are typechecking a co-de Bruijn term the type $\top$ can appear, but because we are forced to discard unused variables immediately, we can never discard a variable that was previously copied, and thus we can never be in a situation where we are taking $\wedge$ of types that involve $\top$. We choose to write the general case anyway because it is needed for many extensions beyond simply-typed, and is arguably also semantically clearer.

The presentation we have given is equivalent to an ordinary `cartesian' $\lambda$-calculus. We can also extend this with rules adapted from the tensor product rules for linear $\lambda$-calculus, and in the presence of co-de Bruijn covers and thinnings it behaves as a cartesian product. However, the standard cartesian product rules from cartesian $\lambda$-calculus, defined by projections, cannot be typechecked in this way (at least without a more advanced notion of subtyping). In the following sections we will use System L to lift this restriction.

\section{System L, the positive fragment}

System L, also known as $\mumut$-calculus, is a term calculus presentation of classical sequent calculus with a rich and well-behaved operational semantics \cite{curien2000duality,munch2013syntax,downen2017sequent,ostermann2022introduction,binder2024grokking}. Unlike $\lambda$-calculus which usually has a single class of terms, System L has three, known as \emph{terms}, \emph{coterms} and \emph{commands}. Each of them has two variable contexts, which we will call the \emph{input context} and \emph{output context}. These should not be confused with inputs and outputs of typecheckers; sometimes they will match and sometimes they will conflict. Terms have a focussed output, coterms have a focussed input, and commands have no focus. It is possible to think of System L terms as somewhat similar to $\lambda$-calculus terms, while coterms are somewhat like first class continuations, and commands describe calling a continuation with a value.

In this section we consider the positive fragment of System L, then over the next few sections we will extend it to full polarised System L.

In this section and for the remainder of the paper we choose to omit writing covers and thinnings in our rules, leaving the scoping information implicit. Importantly, in everything that follows we can choose to either omit or include the discarding and copying rules for thinnings and covers, and varying this gives us either linear or ordinary cartesian System L. Crucially, all of our typechecking rules are agnostic to this choice.

The scoping rules for the positive fragment of system L are given in figure \ref{fig:positive-L-scoping}. System L has two binders: $\mu$ forms a term by binding an output of a command and bringing it into focus, and $\tilde\mu$ forms a coterm by binding an input of a command and bringing it into focus. The only rule for forming a command is \emph{cut}, which assigns the focussed output of a term to the focussed input of a coterm. We write a cut using the syntax $\left< e \mid p \right>$, where $e$ is a term and $p$ is a coterm. (The variables are mnemonic for \emph{expression} and \emph{pattern}, for reasons that will be explained in the next section.)

\begin{figure}[h!]
	\begin{framed}
		\[ \frac{\Gamma_1; \Delta_1 \vdash e\ \term^+ \qquad \Gamma_2; \Delta_2 \vdash p\ \coterm^+}{\Gamma_1, \Gamma_2; \Delta_1; \Delta_2 \vdash \left< e \mid p \right> \cmd^+} \]
		\[ \frac{}{x; \cdot \vdash x\ \term^+} \qquad\qquad \frac{\Gamma; \Delta, x \vdash c\ \cmd^+}{\Gamma; \Delta \vdash \mu x . c\ \term^+} \]
		\[ \frac{}{\cdot\ ; x \vdash x\ \coterm^+} \qquad\qquad \frac{\Gamma, x; \Delta \vdash c\ \cmd^+}{\Gamma; \Delta \vdash \tilde\mu x . c\ \coterm^+} \]
	\end{framed}
	\caption{Scoping rules for positive system L}
	\label{fig:positive-L-scoping}
\end{figure}

We now come to the bidirectional typing rules, and we arrive at the key contribution of this paper: the bidirectional mode coincides with the existing term/coterm/command distinction of System L. Specifically, terms are checkable, coterms are synthesisable, and for commands the distinction does not make sense.

For our notation for typing rules, we continue to generalise Mc Bride's `ni' convention rather than a more standard notation. The fundamental principle is that typechecker inputs always go to the left and typechecker outputs always go to the right, maintaining an invariant that the \emph{information flow in the typechecker} always flows left-to-right. In essence this means we are preferring the perspective of the typechecker to the perspective of the logic.

To accommodate this, we replace turnstiles with the symbols $\Rightarrow$ and $\Leftarrow$ pointing from the \emph{logical} input context to the \emph{logical} output context. In the positive fragment that we are considering in this section, the input context is always a typechecker output and the output context is always a typechecker input, and thus all of our turnstile arrows will point to the left. In the negative fragment this is reversed.

\begin{tabular}{|c|c|c|c|}
	\hline
	Positive System L & Typechecker inputs & Typechecker outputs & Notation \\
	\hline
	Terms & scoped input context, & typed input context & $\inputs\Delta \Leftarrow \inputs A \ni \inputs e \Leftarrow \outputs\Gamma$ \\
	& typed output context, & & \\
	& type, term & & \\
	\hline
	Coterms & scoped input context, & type, typed input context & $\inputs\Delta \Leftarrow \inputs p \in \outputs A \Leftarrow \outputs\Gamma$ \\
	& typed output context, & & \\
	& coterm & & \\
	\hline
	Commands & scoped input context, & typed input context & $\inputs\Delta \Leftarrow \inputs c \Leftarrow \outputs\Gamma$ \\
	& typed output context, & & \\
	& command & & \\
	\hline
\end{tabular}

\begin{itemize}
	\item For a command, which is necessarily a cut, we first synthesise the type of the coterm and then check the term against it:
	\[ \frac{\inputs{\Delta_2} \Leftarrow \inputs p \in \outputs A \Leftarrow \outputs{\Gamma_2} \qquad \inputs{\Delta_1} \Leftarrow \inputs A \ni \inputs e \Leftarrow \outputs{\Gamma_1}}{\inputs{\Delta_1}, \inputs{\Delta_2} \Leftarrow \inputs{\left< e \mid p \right>} \Leftarrow \outputs{\Gamma_1}, \outputs{\Gamma_2}} (\mathrm{cut}^+) \]

	\item To check a variable term, we return the corresponding singleton synthesisable context:
	\[ \frac{}{\inputs\cdot \Leftarrow \inputs A \ni \inputs x \Leftarrow \inputs x : \outputs A} (\mathrm{var}^+) \]

	\item To check a $\mu$-binder, we check the command after extending the checkable context with the type of the bound variable:
	\[ \frac{\inputs\Delta, \inputs x : \inputs A \Leftarrow \inputs c \Leftarrow \outputs\Gamma}{\inputs\Delta \Leftarrow \inputs A \ni \inputs{\mu x . c} \Leftarrow \outputs\Gamma} (\mu^+) \]

	\item To synthesise a variable coterm, we look it up in the checkable context:
	\[ \frac{}{\inputs x : \inputs A \Leftarrow \inputs x \in \outputs A \Leftarrow \outputs\cdot} (\mathrm{covar}^+) \]

	\item To synthesise a $\tilde\mu$-binder, we look up the type of the bound variable in the checkable of the command:
	\[ \frac{\inputs\Delta \Leftarrow \inputs c \Leftarrow \outputs\Gamma, \inputs x : \outputs A}{\inputs\Delta \Leftarrow \inputs{\tilde\mu x . c} \in \outputs A \Leftarrow \outputs\Gamma} (\tilde\mu^+) \]
\end{itemize}

One thing worth noting is that the two variable rules flip the direction of information flow: variable terms are checkable but the variable lives in the synthesisable context, where it must have been bound by a synthesisable connective such as $\tilde\mu$; whereas variable coterms are synthesisable but the variable lives in the checkable context, where it must have been bound by a checkable connective such as $\mu$. This is characteristic of cocontextual typechecking, but is only clearly revealed in the setting of System L where we have both checkable and synthesisable contexts.

To this language we add positive products and coproducts (the latter were notably absent in cocontextual $\lambda$-calculus), whose scoping rules are given in figure \ref{fig:positive-L-scoping-2}. Positive products behave either as linear tensor products or cartesian products, depending on whether we use linear or cartesian scoping. In each case we have  introduction rules which are terms, and ``match'' rules which are coterms that bind commands. The only subtlety is that when we match on a coproduct, both branches must consume every linear variable in the input context, but can produce different output contexts.

Notably, typechecking the combination of a positive product and a coproduct is difficult in both ordinary and cocontextual bidirectional $\lambda$-calculi. Ordinary (contextual) $\lambda$-calculus works well when the only checkable binder is $\lambda$ and struggles with other binders such as the positive product elimination rule, while the cocontextual $\lambda$-calculus overcomes this difficulty but instead forces coproduct introductions to be synthesisable which means that injections require an annotation. We are able to overcome this because System L is bicontextual.

Although technically all logical rules of System L (ie. those corresponding to connectives in the type language) are introduction rules, we adopt the terminology that rules that produce a positive term will be called \emph{introduction rules} and rules that produce a positive coterm will be called \emph{match rules}. These correspond to right and left introduction rules of sequent calculus. The hypotheses of match rules are always commands (possibly none, in the case of nullary coproducts), and the match rule always binds some variables in the (logical) \emph{input} contexts of those commands.

\begin{figure}[h!]
	\begin{framed}
		\[ \frac{\Gamma_1; \Delta_1 \vdash e_1\ \term \qquad \Gamma_2; \Delta_2 \vdash e_2\ \term}{\Gamma_1, \Gamma_2; \Delta_1, \Delta_2 \vdash \left< e_1, e_2 \right>\ \term} \qquad\qquad \frac{\Gamma, x, y; \Delta \vdash c\ \cmd}{\Gamma; \Delta \vdash \match\ \{ \left< x, y \right> \Rightarrow c \}\ \coterm} \]
		\[ \frac{\Gamma; \Delta \vdash e\ \term}{\Gamma; \Delta \vdash \iota_L e\ \term} \qquad\qquad \frac{\Gamma; \Delta \vdash e\ \term}{\Gamma; \Delta \vdash \iota_R e\ \term} \qquad\qquad \frac{\Gamma, x; \Delta_1 \vdash c\ \cmd \qquad\Gamma, y; \Delta_2 \vdash d\ \cmd}{\Gamma; \Delta_1, \Delta_2 \vdash \match\ \{ \iota_L x \Rightarrow c; \iota_R y \Rightarrow d \}\ \coterm} \]
	\end{framed}
	\caption{Scoping rules for tensor products and coproducts in positive linear system L}
	\label{fig:positive-L-scoping-2}
\end{figure}

\begin{itemize}
	\item To check the type of a positive product we check the two subterms:
	\[ \frac{\inputs{\Delta_1} \Leftarrow \inputs A \ni \inputs{e_1} \Leftarrow \outputs{\Gamma_1} \qquad \inputs{\Delta_2} \Leftarrow \inputs B \ni \inputs{e_2} \Leftarrow \outputs{\Gamma_2}}{\inputs{\Delta_1}, \inputs{\Delta_2} \Leftarrow \inputs{A \otimes B} \ni \inputs{\left< e_1, e_2 \right>} \Leftarrow \outputs{\Gamma_1}, \outputs{\Gamma_2}} (\otimes\mathrm{-intro}) \]

	\item To synthesise the type of a positive product match, we check the command and lookup the bound variables in the synthesised context:
	\[ \frac{\inputs\Delta \Leftarrow \inputs c \Leftarrow \outputs\Gamma, \inputs x : \outputs A, \inputs y : \outputs B}{\inputs\Delta \Leftarrow \inputs{\match\ \{ \left< x, y \right> \Rightarrow c \}} \in \outputs{A \otimes B} \Leftarrow \outputs\Gamma} (\otimes\mathrm{-match}) \]

	\item To check the type of a coproduct injection we check the subterm against the appropriate part of the type:
	\[ \frac{\inputs \Delta \Leftarrow \inputs A \ni \inputs e \Leftarrow \outputs\Gamma}{\inputs\Delta \Leftarrow \inputs{A + B} \ni \inputs{\iota_L e} \Leftarrow \outputs\Gamma} (+\mathrm{-intro-L}) \qquad\qquad \frac{\inputs\Delta \Leftarrow \inputs B \ni \inputs e \Leftarrow \outputs\Gamma}{\inputs \Delta \Leftarrow \inputs{A + B} \ni \inputs{\iota_R e} \Leftarrow \outputs\Gamma} (+\mathrm{-intro-R}) \]

	\item So synthesise the type of a coproduct pattern, we check the two commands and lookup the bound variables in the synthesised contexts:
	\[ \frac{\inputs{\Delta_1} \Leftarrow \inputs c \Leftarrow \outputs\Gamma, \inputs x : \outputs A \qquad \inputs{\Delta_2} \Leftarrow \inputs d \Leftarrow \outputs\Gamma, \inputs y : \outputs A}{\inputs{\Delta_1}, \inputs{\Delta_2} \Leftarrow \inputs{\match\ \{ \iota_L x \Rightarrow c; \iota_R y \Rightarrow d \}} \in \outputs{A + B} \Leftarrow \outputs\Gamma} (+\mathrm{-match}) \]
\end{itemize}

\section{The negative fragment}

We now introduce the negative fragment of System L in isolation; the two fragments will be brought together into the full polarised calculus in the next section. All of the rules of the negative fragment are syntactically dual to rules of the positive fragment, but sometimes in counterintuitive ways. We deviate from standard System L terminology, which identifies positive and negative terms as being \emph{producers} and identifies positive and negative coterms as being \emph{consumers}. Instead, we identify positive terms and negative coterms as being \emph{(co)expressions}, and negative terms and positive coterms as being \emph{(co)patterns}. This aligns better with the syntactic symmetries of polarised System L. Thus in the negative fragment we refer to terms as $p$ and coterms as $e$.

Whereas in the positive fragment the logical input context was always synthesisable and the logical output context always checkable, in the negative fragment this reverses, so the logical input context is always checkable and the logical output context is always synthesisable. Thus in the negative fragment we write our turnstiles as $\Rightarrow$, indicating the logical information flow is now also left-to-right.

\begin{tabular}{|c|c|c|c|}
	\hline
	Negative System L & Typechecker inputs & Typechecker outputs & Notation \\
	\hline
	Terms & typed input context, & type, typed output context & $\inputs\Gamma \Rightarrow \inputs p \in \outputs A \Rightarrow \outputs\Delta$ \\
	& scoped output context, term & & \\
	\hline
	Coterms & typed input context, & typed output context & $\inputs\Gamma \Rightarrow \inputs A \ni \inputs e \Rightarrow \outputs\Delta$ \\
	& scoped output context, & & \\
	& type, coterm & & \\
	\hline
	Commands & typed left context, & typed output context & $\inputs\Gamma \Rightarrow \inputs c \Rightarrow \outputs\Delta$ \\
	& scoped output context, command & & \\
	\hline
\end{tabular}


We deviate from the standard notation for cuts in the negative fragments by writing $\left< e \mid p \right>$ where $e$ is the coexpression (which is now a coterm) and $p$ is the copattern (which is now a term). In contrast, usually cuts are always written with the term on the left and the coterm on the right. Similarly to positive fragment, typechecking a cut proceedings by first synthesising the type of the copattern and checking it against the coexpression:
\[ \frac{\inputs{\Gamma_2} \Rightarrow \inputs p \in \outputs A \Rightarrow \outputs{\Delta_2} \qquad \inputs{\Gamma_1} \Rightarrow \inputs A \ni \inputs e \Rightarrow \outputs{\Delta_1}}{\inputs{\Gamma_1}, \inputs{\Gamma_2} \Rightarrow \inputs{\left< e \mid p \right>} \Rightarrow \outputs{\Delta_1}, \outputs{\Delta_2}} (\mathrm{cut}^-) \]

The other structural rules of the negative fragment are formally dual to the corresponding rules in the positive fragment:
\[ \frac{}{\inputs x : \inputs A \Rightarrow \inputs x \in \outputs A \Rightarrow \outputs\cdot} (\mathrm{var}^-) \qquad\qquad \frac{}{\inputs\cdot \Rightarrow \inputs A \ni \inputs x \Rightarrow \inputs x : \outputs A} (\mathrm{covar}^-) \]
\[ \frac{\inputs\Gamma \Rightarrow \inputs c \Rightarrow \outputs\Delta, \inputs x : \outputs A}{\inputs\Gamma \Rightarrow \inputs{\mu x . c} \in \outputs A \Rightarrow \outputs\Delta} (\mu^-) \qquad\qquad \frac{\inputs\Gamma, \inputs x : \inputs A \Rightarrow \inputs c \Rightarrow \outputs\Delta}{\inputs\Gamma \Rightarrow \inputs A \ni \inputs{\tilde\mu x . c} \Rightarrow \outputs\Delta} (\tilde\mu^-) \]

In the negative fragment, we refer to logical rules that produce a term as \emph{comatch} rules, and logical rules that produce a coterm as \emph{cointroduction rules}. Similarly to match rules, comatch rules always have commands as their hypotheses and always bind variables in their (logical) \emph{output} contexts.

The cointroduction and comatch rules for $\parr$ are formally dual to the introduction and match rules for $\otimes$:
\[ \frac{\inputs{\Gamma_1} \Rightarrow \inputs A \ni \inputs{e_1} \Rightarrow \outputs{\Delta_1} \qquad \inputs{\Gamma_2} \Rightarrow \inputs B \ni \inputs{e_2} \Rightarrow \outputs{\Delta_2}}{\inputs{\Gamma_1}, \inputs{\Gamma_2} \Rightarrow \inputs{A \parr B} \ni \inputs{[e_1, e_2]} \Rightarrow \outputs{\Delta_1}, \outputs{\Delta_2}} (\parr\mathrm{-cointro}) \]
\[ \frac{\inputs\Gamma \Rightarrow \inputs c \Rightarrow \outputs\Delta, \inputs x : \outputs A, \inputs y : \outputs B}{\inputs\Gamma \Rightarrow \inputs{\comatch\ \{ c \Rightarrow [x, y] \}} \in \outputs{A \parr B} \Rightarrow \outputs\Delta} (\parr\mathrm{-comatch}) \]
and the cointroduction and comatch rules for $\times$ are formally dual to the introduction and match rules for $+$:
\[ \frac{\inputs\Gamma \Rightarrow \inputs A \ni \inputs e \Rightarrow \outputs\Delta}{\inputs\Gamma \Rightarrow \inputs{A \times B} \ni \inputs{\pi_L e} \Rightarrow \outputs\Delta} (\times\mathrm{-cointro-L}) \qquad\qquad \frac{\inputs\Gamma \Rightarrow \inputs B \ni \inputs e \Rightarrow \outputs\Delta}{\inputs\Gamma \Rightarrow \inputs{A \times B} \ni \inputs{\pi_R e} \Rightarrow \outputs\Delta} (\times\mathrm{-cointro-R}) \]
\[ \frac{\inputs{\Gamma_1} \Rightarrow \inputs c \Rightarrow \outputs\Delta, \inputs x : \outputs A \qquad \inputs{\Gamma_2} \Rightarrow \inputs d \Rightarrow \outputs\Delta, \inputs y : \outputs B}{\inputs{\Gamma_1}, \inputs{\Gamma_2} \Rightarrow \inputs{\comatch\ \{ c \Rightarrow \pi_L x; d \Rightarrow \pi_R y \}} \in \outputs{A \times B} \Rightarrow \outputs\Delta} (\times\mathrm{-comatch}) \]

\section{Full polarised System L}


We now merge the positive and negative fragments of System L into a unified polarised calculus, adding rules for negations and shifts to those already given. In the previous section we emphasised the syntactic symmetry between the positive and negative fragments, and it is possible to lean into this symmetry and build our syntax around it in a way that halves the total number of judgements. In this paper we have chosen not to do this, favouring clarity over brevity in our syntax. 
However, the bidirectional typechecking modes do align in exactly this way: (co)expressions are always checkable and (co)patterns are always synthesisable.

We need to modify the existing rules by having positive judgements carry around negative contexts and vice versa. We will continue to use our arrow notation for turnstiles, so for example the judgement of a positive term will now be written $\inputs{\Gamma^-}; \inputs{\Delta^+} \Leftarrow \inputs{A^+} \ni \inputs{e^+} \Leftarrow \outputs{\Gamma^+}; \outputs{\Delta^-}$. Although it appears that the logical information flow from the negative input context $\Gamma^-$ to the negative output context $\Delta^-$ now goes in the wrong direction, we justify this by viewing the judgement from the perspective of its `native' fragment.

A further subtlety is that commands are \emph{not} polarised: we have a single command judgement that subsumes both positive and negative commands, with a positive cut rule that assigns an expression to a pattern, and a negative cut rule that assigns a coexpression to a copattern. Unfortunately this unity of commands breaks our syntax for turnstiles; we instead will use the symbol $\rightleftharpoons$ for the turnstiles of commands.



\hspace{-2cm}\begin{tabular}{|c|c|c|c|}
	\hline
	Full polarised & Typechecker inputs & Typechecker outputs & Notation \\
	System L & & & \\
	\hline
	(Co)expressions & scoped positive input context, & typed positive input context, & $\inputs{\Gamma^-}; \inputs{\Delta^+} \Leftarrow \inputs{A^+} \ni \inputs{e^+} \Leftarrow \outputs{\Gamma^+}; \outputs{\Delta^-}$ \\
	(positive terms, & typed negative input context, & typed negative output context & $\inputs{\Gamma^-}; \inputs{\Delta^+} \Rightarrow \inputs{A^-} \ni \inputs{e^-} \Rightarrow \outputs{\Gamma^+}; \outputs{\Delta^-}$ \\
	negative coterms) & typed positive output context, & & \\
	& scoped negative output context, & & \\
	& type, expression & & \\
	\hline
	(Co)patterns & scoped positive input context, & type, & $\inputs{\Gamma^-}; \inputs{\Delta^+} \Leftarrow \inputs{p^+} \in \outputs{A^+} \Leftarrow \outputs{\Gamma^+}; \outputs{\Delta^-}$ \\
	(positive coterms, & typed negative input context, & typed positive input context, & $\inputs{\Gamma^-}; \inputs{\Delta^+} \Rightarrow \inputs{p^-} \in \outputs{A^-} \Rightarrow \outputs{\Gamma^+}; \outputs{\Delta^-}$ \\
	negative terms) & typed positive output context, & typed negative output context & \\
	& scoped negative output context, & & \\
	& (co)pattern & & \\
	\hline
	Commands & scoped positive input context, & typed positive input context, & $\inputs{\Gamma^-}; \inputs{\Delta^+} \rightleftharpoons \inputs c \rightleftharpoons \outputs{\Gamma^+}; \outputs{\Delta^-}$ \\
	& typed negative input context, & typed negative output context & \\
	& typed positive output context, & & \\
	& scoped negative output context, & & \\
	& command & & \\
	\hline
\end{tabular}


We extend our type language with a pair of negations and a pair of shifts, which go between positive and negative types. The full type language of polarised System L is now described by:
\[ A^+ ::= I \mid A^+ \otimes B^+ \mid 0 \mid A^+ + B^+ \mid\ \sim\! A^- \mid\ \downarrow\! A^- \]
\[ A^- ::= \bot \mid A^- \parr B^- \mid 1 \mid A^- \times B^- \mid \neg A^+ \mid\ \uparrow\! A^+ \]

The ways that negations and shifts convert between the 4 judgement sorts are described by the following diagram:

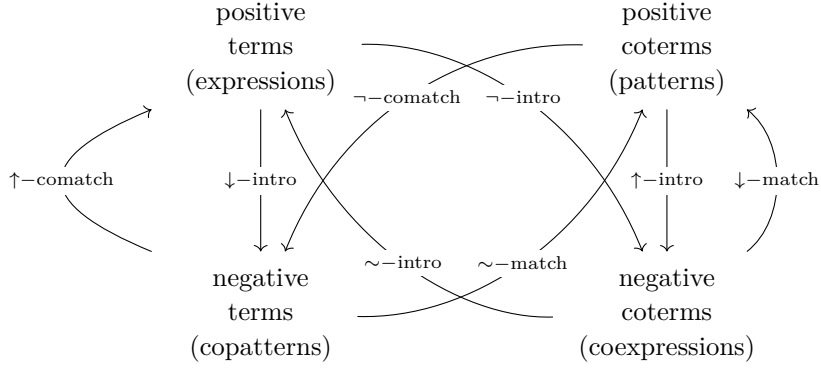
\[\begin{tikzcd}[scale=1.5]
	\begin{array}{c} \mathrm{positive} \\ \mathrm{terms} \\ \mathrm{(expressions)} \end{array} &&& \begin{array}{c} \mathrm{positive} \\ \mathrm{coterms} \\ \mathrm{(patterns)} \end{array} \\
	\\
	\\
	\begin{array}{c} \mathrm{negative} \\ \mathrm{terms} \\ \mathrm{(copatterns)} \end{array} &&& \begin{array}{c} \mathrm{negative} \\ \mathrm{coterms} \\ \mathrm{(coexpressions)} \end{array}
	\arrow["{\downarrow\mathrm{-intro}}"{description}, from=1-1, to=4-1]
	\arrow["{\neg\mathrm{-intro}}"{description}, curve={height=-30pt}, from=1-1, to=4-4]
	\arrow["{\neg\mathrm{-comatch}}"{description}, curve={height=30pt}, from=1-4, to=4-1]
	\arrow["{\uparrow\mathrm{-intro}}"{description}, from=1-4, to=4-4]
	\arrow["{\uparrow\mathrm{-comatch}}"{description}, shift left=3, curve={height=-30pt}, from=4-1, to=1-1]
	\arrow["{\sim\mathrm{-match}}"{description}, curve={height=30pt}, from=4-1, to=1-4]
	\arrow["{\sim\mathrm{-intro}}"{description}, curve={height=-30pt}, from=4-4, to=1-1]
	\arrow["{\downarrow\mathrm{-match}}"{description}, curve={height=30pt}, from=4-4, to=1-4]
\end{tikzcd}\]

\begin{itemize}
	\item The introduction rule for positive negation takes a negative coterm to a positive term (and thus a coexpression to an expression), so it reverses the chirality while preserving the typechecking mode:
	\[ \frac{\inputs{\Gamma^-}; \inputs{\Delta^+} \Rightarrow \inputs{A^-} \ni \inputs{e^-} \Rightarrow \outputs{\Gamma^+}; \outputs{\Delta^-}}{\inputs{\Gamma^-}; \inputs{\Delta^+} \Leftarrow \inputs{\sim\! A^-} \ni \inputs{\sim\! (e^-)} \Leftarrow \outputs{\Gamma^+}; \outputs{\Delta^-}} (\sim\mathrm{-intro}) \]

	\item The match rule for positive negation takes a command with a free negative output variable to a positive coterm (pattern), using the synthesised type of the free variable to synthesise the type of the coterm:
	\[ \frac{\inputs{\Gamma^-}; \inputs{\Delta^+} \rightleftharpoons \inputs c \rightleftharpoons \outputs{\Gamma^+}; \outputs{\Delta^-}, \inputs x : \outputs{A^-}}{\inputs{\Gamma^-}; \inputs{\Delta^+} \Leftarrow \inputs{\mathrm{match} \{ \sim\! x \Rightarrow c \}} \in \outputs{\sim\! A^-} \Leftarrow \outputs{\Gamma^+}; \outputs{\Delta^-}} (\sim\mathrm{-match}) \]

	\item The cointroduction rule for negative negation takes a positive term to a negative coterm (and thus an expression to a coexpression), the opposite of the positive negation introduction rule:
	\[ \frac{\inputs{\Gamma^-}; \inputs{\Delta^+} \Leftarrow \inputs{A^+} \ni \inputs{e^+} \Leftarrow \outputs{\Gamma^+}; \outputs{\Delta^-}}{\inputs{\Gamma^-}; \inputs{\Delta^+} \Rightarrow \inputs{\neg A^+} \ni \inputs{\neg (e^+)} \Rightarrow \outputs{\Gamma^+}; \outputs{\Delta^-}} (\neg\mathrm{-cointro}) \]

	\item The comatch rule for negative negation takes a command with a free positive input variable to a negative term (copattern), using the synthesised type of the free variable to synthesise the type of the term:
	\[ \frac{\inputs{\Gamma^-}; \inputs{\Delta^+} \rightleftharpoons \inputs c \rightleftharpoons \outputs{\Gamma^+}, \inputs x : \outputs{A^+}; \outputs{\Delta^-}}{\inputs{\Gamma^-}; \inputs{\Delta^+} \Rightarrow \inputs{\mathrm{comatch} \{ c \Rightarrow \neg x \}} \in \outputs{\neg A^+} \Rightarrow \outputs{\Gamma^+}; \outputs{\Delta^-}} (\neg\mathrm{-comatch}) \]

	\item The introduction rule for downshift takes a negative term (copattern) to a positive term (expression), switching the bidirectional mode from synthesisable to checkable by checking a subtype:
	\[ \frac{\inputs{\Gamma^-}; \inputs{\Delta^+} \Rightarrow \inputs{p^-} \in \outputs{B^-} \Rightarrow \outputs{\Gamma^+}; \outputs{\Delta^-} \qquad \inputs{A^-} \sqsubseteq \inputs{B^-}}{\inputs{\Gamma^-}; \inputs{\Delta^+} \Leftarrow \inputs{\downarrow\! A^-} \ni \inputs{\downarrow\! p^-} \Leftarrow \outputs{\Gamma^+}; \outputs{\Delta^-}} (\downarrow\mathrm{-intro}) \]

	\item The match rule for downshift takes a command with a free negative input variable to a positive coterm (pattern), switching the bidirectional mode from checkable to synthesisable and thus requiring an annotation:
	\[ \frac{\inputs{\Gamma^-}, \inputs x : \inputs{A^-}; \inputs{\Delta^+} \rightleftharpoons \inputs c \rightleftharpoons \outputs{\Gamma^+}; \outputs{\Delta^-}}{\inputs{\Gamma^-}; \inputs{\Delta^+} \Leftarrow \inputs{\mathrm{match} \{ \downarrow\! (x :: A^-) \Rightarrow c \}} \in \outputs{\downarrow\! A^-} \Leftarrow \outputs{\Gamma^+}; \outputs{\Delta^-}} (\downarrow\mathrm{-match}) \]

	\item The cointroduction rule for upshift takes a positive coterm (pattern) to a negative coterm (coexpression), switching the bidirectional mode from synthesisable to checkable by checking a subtype:
	\[ \frac{\inputs{\Gamma^-}; \inputs{\Delta^+} \Leftarrow \inputs{p^+} \in \outputs{B^+} \Leftarrow \outputs{\Gamma^+}; \outputs{\Delta^-} \qquad \inputs{B^+} \sqsubseteq \inputs{A^+}}{\inputs{\Gamma^-}; \inputs{\Delta^+} \Rightarrow \inputs{\uparrow\! A^+} \ni \inputs{\uparrow\! p^+} \Rightarrow \outputs{\Gamma^+}; \outputs{\Delta^-}} (\uparrow\mathrm{-cointro}) \]

	\item The comatch rule for upshift takes a command with a free positive output variable to a negative term (copattern), switching the bidirectional mode from checkable to synthesisable and thus requiring an annotation:
	\[ \frac{\inputs{\Gamma^-}; \inputs{\Delta^+}, \inputs x : \inputs{A^+} \rightleftharpoons \inputs c \rightleftharpoons \outputs{\Gamma^+}; \outputs{\Delta^-}}{\inputs{\Gamma^-}; \inputs{\Delta^+} \Rightarrow \inputs{\mathrm{comatch} \{ c \Rightarrow\ \uparrow\! (x :: A^+) \}} \in \outputs{\uparrow\! A^+} \Rightarrow \outputs{\Gamma^+}; \outputs{\Delta^-}} (\uparrow\mathrm{-comatch}) \]
\end{itemize}

Since all of the typing rules throughout this paper have been interleaved with explanations, we will now repeat the full definition of bicontextual polarised System L, given in figures \ref{fig:full-L-positive}, \ref{fig:full-L-negative} and \ref{fig:full-L-mixed}. To save space we finally take full advantage of System L's symmetry by merging contexts. We use $X$ (a capital $\chi$) to refer to a merged checkable context $\Gamma^-; \Delta^+$, and $\Sigma$ to refer to a merged synthesisable context $\Gamma^+; \Delta^-$. When binding a variable in a context, we will use superscripts to indicate its polarity. Thus for example the notation $\Sigma, x^+ : A^+$ means $\Gamma^+, x^+ : A^+; \Delta^-$.

\begin{figure}
	\begin{framed}
		\[ \frac{}{\inputs\cdot \Leftarrow \inputs{A^+} \ni \inputs{x^+} \Leftarrow \inputs{x^+} : \outputs{A^+}} (\mathrm{var}^+) \qquad\qquad \frac{}{\inputs{x^+} : \inputs{A^+} \Leftarrow \inputs{x^+} \in \outputs{A^+} \Leftarrow \outputs\cdot} (\mathrm{covar}^+) \]
		\[ \frac{\inputs X, \inputs{x^+} : \inputs{A^+} \rightleftharpoons \inputs c \rightleftharpoons \outputs\Sigma}{\inputs X \Leftarrow \inputs{A^+} \ni \inputs{\mu^+ x^+ . c} \Leftarrow \outputs\Sigma} (\mu^+) \qquad\qquad \frac{\inputs X \rightleftharpoons \inputs c \rightleftharpoons \outputs\Sigma, \inputs{x^+} : \outputs{A^+}}{\inputs X \Leftarrow \inputs{\tilde\mu^+ x^+ . c} \in \outputs{A^+} \Leftarrow \outputs\Sigma} (\tilde\mu^+) \]
		\[ \frac{}{\inputs\cdot \Leftarrow \inputs I \ni \inputs{\left< \right>} \Leftarrow \outputs\cdot} (I\mathrm{-intro}) \qquad\qquad \frac{\inputs X \rightleftharpoons \inputs c \rightleftharpoons \outputs\Sigma}{\inputs X \Leftarrow \inputs{\mathrm{match} \{ \left< \right> \Rightarrow c \}} \in \outputs I \Leftarrow \outputs\Sigma} (I\mathrm{-match}) \]
		\[ \frac{\inputs{X_1} \Leftarrow \inputs{A^+} \ni \inputs{e_1^+} \Leftarrow \outputs{\Sigma_1} \qquad \inputs{X_2} \Leftarrow \inputs{B^+} \ni \inputs{e_2^+} \Leftarrow \outputs{\Sigma_2}}{\inputs{X_1}, \inputs{X_2} \Leftarrow \inputs{A^+ \otimes B^+} \ni \inputs{\left< e_1^+, e_2^+\right>} \Leftarrow \outputs{\Sigma_1}, \outputs{\Sigma_2}} (\otimes\mathrm{-intro}) \]
		\[ \frac{\inputs X \rightleftharpoons \inputs c \rightleftharpoons \outputs\Sigma, \inputs{x^+} : \outputs{A^+}, \inputs{y^+} : \outputs{B^+}}{\inputs X \Leftarrow \inputs{\match\ \{ \left< x^+, y^+ \right> \Rightarrow c \}} \in \outputs{A^+ \otimes B^+} \Leftarrow \outputs\Sigma} (\otimes\mathrm{-match}) \]
		\[ \frac{\inputs X \Leftarrow \inputs{A^+} \ni \inputs{e^+} \Leftarrow \outputs\Sigma}{\inputs X \Leftarrow \inputs{A^+ + B^+} \ni \inputs{\iota_L e^+} \Leftarrow \outputs\Sigma} (+\mathrm{-intro-L}) \qquad\qquad \frac{\inputs X \Leftarrow \inputs{B^+} \ni \inputs{e^+} \Leftarrow \outputs\Sigma}{\inputs X \Leftarrow \inputs{A^+ + B^+} \ni \inputs{\iota_R e^+} \Leftarrow \outputs\Sigma} (+\mathrm{-intro-R}) \]
		\[ \frac{}{\inputs X \Leftarrow \inputs{\mathrm{match} \{ \}} \in \outputs 0 \Leftarrow \outputs \Sigma} (0\mathrm{-match}) \]
		\[ \frac{\inputs{X_1} \rightleftharpoons \inputs c \rightleftharpoons \outputs\Sigma, \inputs{x^+} : \outputs{A^+} \qquad \inputs{X_2} \rightleftharpoons \inputs d \rightleftharpoons \outputs\Sigma, \inputs{y^+} : \outputs{B^+}}{\inputs{X_1}, \inputs{X_2} \Leftarrow \inputs{\match\ \{ \iota_L x^+ \Rightarrow c; \iota_R y^+ \Rightarrow d \}} \in \outputs{A^+ + B^+} \Leftarrow \outputs\Sigma} (+\mathrm{-match}) \]
	\end{framed}
	\caption{Full bicontextual polarised System L, positive rules}
	\label{fig:full-L-positive}
\end{figure}

\begin{figure}
	\begin{framed}
		Negative fragment:
		\[ \frac{}{\inputs{x^-} : \inputs{A^-} \Rightarrow \inputs{x^-} \in \outputs{A^-} \Rightarrow \outputs\cdot} (\mathrm{var}^-) \qquad\qquad \frac{}{\inputs\cdot \Rightarrow \inputs{A^-} \ni \inputs{x^-} \Rightarrow \inputs{x^-} : \outputs{A^-}} (\mathrm{covar}^-) \]
		\[ \frac{\inputs X \rightleftharpoons \inputs c \rightleftharpoons \outputs\Sigma, \inputs{x^-} : \outputs{A^-}}{\inputs X \Rightarrow \inputs{\mu^- x^- . c} \in \outputs{A^-} \Rightarrow \outputs\Sigma} (\mu^-) \qquad\qquad \frac{\inputs X, \inputs{x^-} : \inputs{A^-} \rightleftharpoons \inputs c \rightleftharpoons \outputs\Sigma}{\inputs X \Rightarrow \inputs{A^-} \ni \inputs{\tilde\mu^- x^- . c} \Rightarrow \outputs\Sigma} (\tilde\mu^-) \]
		\[ \frac{}{\inputs\cdot \Rightarrow \inputs\bot \ni \inputs{[\ ]} \Rightarrow \outputs\cdot} (\bot\mathrm{-cointro}) \qquad\qquad \frac{\inputs X \rightleftharpoons \inputs c \rightleftharpoons \outputs\Sigma}{\inputs X \Rightarrow \inputs{\mathrm{comatch} \{ c \Rightarrow [\ ] \}} \in \outputs\bot \Rightarrow \outputs\Sigma} (\bot\mathrm{-comatch}) \]
		\[ \frac{\inputs{X_1} \Rightarrow \inputs{A^-} \ni \inputs{e^-_1} \Rightarrow \outputs{\Sigma_1} \qquad \inputs{X_2} \Rightarrow \inputs{B^-} \ni \inputs{e^-_2} \Rightarrow \outputs{\Sigma_2}}{\inputs{X_1}, \inputs{X_2} \Rightarrow \inputs{A^- \parr B^-} \ni \inputs{[e^-_1, e^-_2]} \Rightarrow \outputs{\Sigma_1}, \outputs{\Sigma_2}} (\parr\mathrm{-cointro}) \]
		\[ \frac{\inputs X \rightleftharpoons \inputs c \rightleftharpoons \outputs\Sigma, \inputs{x^-} : \outputs{A^-}, \inputs{y^-} : \outputs{B^-}}{\inputs X \Rightarrow \inputs{\comatch\ \{ c \Rightarrow [x^-, y^-] \}} \in \outputs{A^- \parr B^-} \Rightarrow \outputs\Sigma} (\parr\mathrm{-comatch}) \]
		\[ \frac{}{\inputs X \Rightarrow \inputs{\mathrm{comatch} \{ \}} \in \outputs 0 \Rightarrow \outputs\Sigma} (0\mathrm{-comatch}) \]
		\[ \frac{\inputs X \Rightarrow \inputs{A^-} \ni \inputs{e^-} \Rightarrow \outputs\Sigma}{\inputs X \Rightarrow \inputs{A^- \times B^-} \ni \inputs{\pi_L e^-} \Rightarrow \outputs\Sigma} (\times\mathrm{-cointro-L}) \qquad\qquad \frac{\inputs X \Rightarrow \inputs{B^-} \ni \inputs{e^-} \Rightarrow \outputs\Sigma}{\inputs X \Rightarrow \inputs{A^- \times B^-} \ni \inputs{\pi_R e^-} \Rightarrow \outputs\Sigma} (\times\mathrm{-cointro-R}) \]
		\[ \frac{\inputs{X_1} \rightleftharpoons \inputs c \rightleftharpoons \outputs\Sigma, \inputs{x^-} : \outputs{A^-} \qquad \inputs{X_2} \rightleftharpoons \inputs d \rightleftharpoons \outputs\Sigma, \inputs{y^-} : \outputs{B^-}}{\inputs{X_1}, \inputs{X_2} \Rightarrow \inputs{\comatch\ \{ c \Rightarrow \pi_L x; d \Rightarrow \pi_R y \}} \in \outputs{A \times B} \Rightarrow \outputs\Sigma} (\times\mathrm{-comatch}) \]
	\end{framed}
	\caption{Full bicontextual polarised System L, negative rules}
	\label{fig:full-L-negative}
\end{figure}

\begin{figure}
	\begin{framed}
		Cuts, negations and shifts:
		\[ \frac{\inputs{X_2} \Leftarrow \inputs{p^+} \in \outputs{A^+} \Leftarrow \outputs{\Sigma_2} \qquad \inputs{X_1} \Leftarrow \inputs{A^+} \ni \inputs{e^+} \Leftarrow \outputs{\Sigma_1}}{\inputs{X_1, X_2} \rightleftharpoons \inputs{\left< e^+ \mid p^+ \right>} \rightleftharpoons \outputs{\Sigma_1, \Sigma_2}} (\mathrm{cut}^+) \]
		\[ \frac{\inputs{X_2} \Rightarrow \inputs{p^-} \in \outputs{A^-} \Rightarrow \outputs{\Sigma_2} \qquad \inputs{X_1} \Rightarrow \inputs{A^-} \ni \inputs{e^-} \Rightarrow \outputs{\Sigma_1}}{\inputs{X_1, X_2} \rightleftharpoons \inputs{\left< e^- \mid p^- \right>} \rightleftharpoons \outputs{\Sigma_1, \Sigma_2}} (\mathrm{cut}^-) \]
		\[ \frac{\inputs X \Rightarrow \inputs{A^-} \ni \inputs{e^-} \Rightarrow \outputs\Sigma}{\inputs X \Leftarrow \inputs{\sim\! A^-} \ni \inputs{\sim\! (e^-)} \Leftarrow \outputs\Sigma} (\sim\mathrm{-intro}) \qquad\qquad \frac{\inputs X \rightleftharpoons \inputs c \rightleftharpoons \outputs\Sigma, \inputs{x^-} : \outputs{A^-}}{\inputs X \Leftarrow \inputs{\mathrm{match} \{ \sim\! x \Rightarrow c \}} \in \outputs{\sim\! A^-} \Leftarrow \outputs\Sigma} (\sim\mathrm{-match}) \]
		\[ \frac{\inputs X \Leftarrow \inputs{A^+} \ni \inputs{e^+} \Leftarrow \outputs\Sigma}{\inputs X \Rightarrow \inputs{\neg A^+} \ni \inputs{\neg (e^+)} \Rightarrow \outputs\Sigma} (\neg\mathrm{-cointro}) \qquad\qquad \frac{\inputs X \rightleftharpoons \inputs c \rightleftharpoons \outputs\Sigma, \inputs{x^+} : \outputs{A^+}}{\inputs X \Rightarrow \inputs{\mathrm{comatch} \{ c \Rightarrow \neg x \}} \in \outputs{\neg A^+} \Rightarrow \outputs\Sigma} (\neg\mathrm{-comatch}) \]
		\[ \frac{\inputs X \Rightarrow \inputs{p^-} \in \outputs{B^-} \Rightarrow \outputs\Sigma \qquad \inputs{A^-} \sqsubseteq \inputs{B^-}}{\inputs X \Leftarrow \inputs{\downarrow\! A^-} \ni \inputs{\downarrow\! p^-} \Leftarrow \outputs\Sigma} (\downarrow\mathrm{-intro}) \qquad\qquad \frac{\inputs X, \inputs{x^-} : \inputs{A^-} \rightleftharpoons \inputs c \rightleftharpoons \outputs\Sigma}{\inputs X \Leftarrow \inputs{\mathrm{match} \{ \downarrow\! (x :: A^-) \Rightarrow c \}} \in \outputs{\downarrow\! A^-} \Leftarrow \outputs\Sigma} (\downarrow\mathrm{-match}) \]
		\[ \frac{\inputs X \Leftarrow \inputs{p^+} \in \outputs{B^+} \Leftarrow \outputs\Sigma \qquad \inputs{B^+} \sqsubseteq \inputs{A^+}}{\inputs X \Rightarrow \inputs{\uparrow\! A^+} \ni \inputs{\uparrow\! p^+} \Rightarrow \outputs\Sigma} (\uparrow\mathrm{-cointro}) \qquad\qquad \frac{\inputs X, \inputs{x^+} : \inputs{A^+} \rightleftharpoons \inputs c \rightleftharpoons \outputs\Sigma}{\inputs X \Rightarrow \inputs{\mathrm{comatch} \{ c \Rightarrow\ \uparrow\! (x :: A^+) \}} \in \outputs{\uparrow\! A^+} \Rightarrow \outputs\Sigma} (\uparrow\mathrm{-comatch}) \]
	\end{framed}
	\caption{Full bicontextual polarised System L, mixed rules}
	\label{fig:full-L-mixed}
\end{figure}

\section{System LNL}

In the previous sections we have been implicitly working with both linear and cartesian (ie. ordinary) System L by ambiguously either including or not including the rules for discarding and copying as part of our definitions of thinning and covering. In this section we will show that these can be combined into a full classical linear logic with exponentials. Specifically, we present a linear-nonlinear (LNL) type theory \cite{benton1994mixed} and find that its division into linear and cartesian terms precisely matches the symmetries we have already been considering. We give this linear-nonlinear extension of System L the name System LNL.

LNL calculi generally come with a restriction that linear contexts are omitted in cartesian terms. We find no need for this restriction from considerations of typechecking, and thus we present a more general adjunction calculus in which all judgements have access to all contexts \cite{pruiksma2018adjoint}. We expect that for true LNL it is still necessary to make these restrictions for semantic reasons.

We begin with a fragment of System LNL that admits the linear $!$ modality but not the $?$ modality. To do this, we take the rules for full polarised linear System L, and then we add the rules for discarding in thinnings and copying in covers only for negative input contexts. This gives us weakening and contraction for negative inputs, and has the consequence that negative terms (copatterns) behave in a cartesian way while the others stay linear.

We now need to add the rules for the two adjoint modalities of LNL which give the factorisation of the $!$ comonad. We find that the forgetful functor (left adjoint) from cartesian to linear exactly coincides with our existing downshift modality $\downarrow$, while the cofree functor (right adjoint) from linear to cartesian does not coincide with any connective that was already present. We name the cofree functor $\Uparrow$, so we get a derived comonadic modality defined by $!A =\ \downarrow \Uparrow\! A$. The cointroduction rule for $\Uparrow$ also requires an annotation. This means that the derived rules for $!$ require annotations for both introduction and match, coming respectively from the cointroduction rule for $\Uparrow$ and the match rule for $\downarrow$.

\begin{itemize}
	\item The cointroduction rule for $\Uparrow$ takes a positive term (expression) to a negative term (copattern), switching the bidirectional mode from checkable to synthesisable and thus requiring an annotation:
	\[ \frac{\inputs{\Gamma^-}; \inputs{\Delta^+} \Leftarrow \inputs{A^+} \ni \inputs{e^+} \Leftarrow \outputs{\Gamma^+}; \outputs{\Delta^-}}{\inputs{\Gamma^-}; \inputs{\Delta^+} \Rightarrow \inputs{\Uparrow\! (e^+ :: A^+)} \in \outputs{\Uparrow\! A^+} \Rightarrow \outputs{\Gamma^+}; \outputs{\Delta^-}} \]

	\item The comatch rule for $\Uparrow$ takes a command with a free positive input variable to a negative coterm (coexpression), switching the bidirectional mode from synthesisable to checkable by checking a subtype:
	\[ \frac{\inputs{\Gamma^-}; \inputs{\Delta^+} \rightleftharpoons \inputs c \rightleftharpoons \outputs{\Gamma^+}, \inputs x : \outputs{B^+}; \outputs{\Delta^-} \qquad \inputs{B^+} \sqsubseteq \inputs{A^+}}{\inputs{\Gamma^-}; \inputs{\Delta^+} \Rightarrow \inputs{\Uparrow\! A^+} \ni \inputs{\mathrm{comatch} \{ c \Rightarrow\ \Uparrow\! x \}} \Rightarrow \outputs{\Gamma^+}; \outputs{\Delta^-}} \]
\end{itemize}

We will now restore the symmetry we have broken, adding another modality that allows deriving the $?$ of classical linear logic. This is not possible in standard intuitionistic LNL, but instead we use an embedding of \emph{co-LNL} \cite{eades2017cointuitionistic}. We need to also adding the discarding and copying rules for thinnings and covers in positive output contexts, giving us coweakening and cocontraction for positive outputs, and making positive coterms (patterns) behave in a cocartesian way. Notice that in full System LNL, (co)expressions are always linear while patterns are cocartesian and copatterns are cartesian.

We have one final operator to add, a downshift modality $\Downarrow$ that gives us a derived $?$ monad by $?A =\ \uparrow \Downarrow\! A$. Just as for $!$, the introduction rule for $\Downarrow$ requires an annotation and thus the derived rules for $?$ require annotations for both introduction and matching.

\begin{itemize}
	\item The introduction rule for $\Downarrow$ takes a negative coterm (coexpression) to a positive coterm (pattern), switching the bidirectional mode from checkable to synthesisable and thus requiring an annotation:
	\[ \frac{\inputs{\Gamma^-}; \inputs{\Delta^+} \Rightarrow \inputs{A^-} \ni \inputs{e^-} \Rightarrow \outputs{\Gamma^+}; \outputs{\Delta^-}}{\inputs{\Gamma^-}; \inputs{\Delta^+} \Leftarrow \inputs{\Downarrow\! (e^- :: A^-)} \in \outputs{\Downarrow\! A^-} \Leftarrow \outputs{\Gamma^+}; \outputs{\Delta^-}} \]

	\item The match rule for $\Downarrow$ takes a command with a free negative output variable to a positive term (expression), switching the bidirectional mode from synthesisable to checkable by checking a subtype:
	\[ \frac{\inputs{\Gamma^-}; \inputs{\Delta^+} \rightleftharpoons \inputs c \rightleftharpoons \outputs{\Gamma^+}; \outputs{\Delta^-}, \inputs x : \outputs{B^-} \qquad \inputs{A^-} \sqsubseteq \inputs{B^-}}{\inputs{\Gamma^-}; \inputs{\Delta^+} \Leftarrow \inputs{\Downarrow\! A^-} \ni \inputs{\mathrm{match} \{ \Downarrow\! x \Rightarrow c \}} \Leftarrow \outputs{\Gamma^+}; \outputs{\Delta^-}} \]
\end{itemize}

\begin{figure}
	\caption{System LNL}
\end{figure}

\section{Further work}

\subsection{Embedding of $\lambda$-calculi}

As explained at the end of the introduction, it was a major motivation for us to develop a methdology for designing typecheckers for `ordinary' system such as $\lambda$-calculi, but we have left the details of this for future work. We expect that it will be useful to start from systems that already have some form of polarisation: one conclusion of our work is that bidirectional typechecking is sensitive to polarisation that is present metatheoretically, and therefore gives a new reason to make it explicit.

\subsection{Operational semantics and subject reduction}

System L is known to have an extremely well behaved operational semantics described by CK machines \cite{felleisen_etal_calculus_assignments,curien2000duality}. While System L is already known to satisfy subject reduction (ie. that its operational semantics preserves types) \cite{munch2009focalisation}, we conjecture that by pairing the `natural' (ie. cocontextual) type theory with the `natural' operational semantics, a more parsimonious proof of subject reduction can be obtained, laying a foundation for proving subject reduction in the presence of more advanced type system features.

A standard presentation of polarised System L splits (co)expressions further into values and stacks, and this is crucial to obtain a confluent operational semantics. In this paper this distinction has played no role in typechecking and so we have omitted it, however we believe that it may play a role in systems with subtyping.

\subsection{Subtyping, intersection and union types}

Throughout this paper we have seen that an underspecified subtyping relationship and greatest lower bound operator play a role in bicontextual typechecking, but we have considered only the trivial case where subtyping reduces to equality. This leads us to believe that System L will be a natural setting to study typechecking for richer theories of subtyping and intersection types (and, due to the symmetry of System L, also union types), building on existing work studying those features in System L \cite{tsukada2018intersection,downen2019duality}.

\subsection{Polymorphism and System F}

Although in this paper we have not included a system with type quantification, presentations of polymorphic System L are known \cite{ostermann2022introduction} and our typing discipline extends to these systems. The caveat is that type variable contexts and substitution are a significant technical complication; in practice these are practical issues of implementation rather than theoretical barriers.

Going further, datatype constructors and other polymorphic functions are necessarily forced to live in the negative fragment even when the datatype is strictly positive, because universally quantified types are always negative. We believe this issue can be overcome by a system in which System F-style explicit quantification exists alongside a novel form of let-polymorphism reminiscent of Hindley-Milner type systems.

Interestingly, typechecking a cut between an expression and a pattern that has a free type variable seems to only require a restricted form of unification known as \emph{matching} \cite{spj-scrap-applications} (not to be confused with matching in System L). Our hope is that this will lead to dramatic simplifications in typecheckers for polymorphic type theories.

\subsection{Dependent types}
	
Ultimately, our end goal is to develop a bidirectional dependently-typed extension of polarised System L building on \cite{miquey2019classical,binder2024deriving}, and use it to develop better typecheckers for dependent type theories, where we expect it to give new insights into difficult problems such as higher-order unification and dependent pattern matching.

\bibliography{bidirectional-paper}

\begin{thebibliography}{10}

\bibitem{benton1994mixed}
P~Nick Benton.
\newblock A mixed linear and non-linear logic: Proofs, terms and models.
\newblock In {\em International Workshop on Computer Science Logic}, pages
  121--135. Springer, 1994.

\bibitem{binder2024deriving}
David Binder, Ingo Skupin, Tim S{\"u}berkr{\"u}b, and Klaus Ostermann.
\newblock Deriving dependently-typed oop from first principles.
\newblock {\em Proceedings of the ACM on Programming Languages},
  8(OOPSLA1):983--1009, 2024.

\bibitem{binder2024grokking}
David Binder, Marco Tzschentke, Marius M{\"u}ller, and Klaus Ostermann.
\newblock Grokking the sequent calculus (functional pearl).
\newblock {\em Proceedings of the ACM on Programming Languages},
  8(ICFP):395--425, 2024.

\bibitem{mcbride-everybodys-somewhere}
Conor~Mc Bride.
\newblock Everybody's got to be somewhere.
\newblock In {\em Proceedings of {MSFP} 2018}, volume 275 of {\em EPTCS}, 2018.

\bibitem{mcbride-ni}
Conor~Mc Bride.
\newblock The types who say 'ni'.
\newblock Available at
  \url{https://github.com/pigworker/TypesWhoSayNi/blob/master/tex/TypesWhoSayNi.pdf},
  2019.

\bibitem{mcbride-iso}
Conor~Mc Bride.
\newblock Input < subject > output.
\newblock Available at
  \url{https://github.com/pigworker/TypesWhoSayNi/blob/master/tex/Azer/iso.pdf},
  2025.

\bibitem{curien2000duality}
Pierre-Louis Curien and Hugo Herbelin.
\newblock The duality of computation.
\newblock {\em ACM sigplan notices}, 35(9):233--243, 2000.

\bibitem{downen2017sequent}
Paul Downen.
\newblock {\em Sequent Calculus: A Logic and a Language for Computation and
  Duality}.
\newblock PhD thesis, University of Oregon, 2017.

\bibitem{downen2018beyond}
Paul Downen and Zena~M Ariola.
\newblock Beyond polarity: Towards a multi-discipline intermediate language
  with sharing.
\newblock In {\em Conference on Computer Science Logic (CSL 2018)}, volume 119,
  2018.

\bibitem{downen2019duality}
Paul Downen, Zena~M Ariola, and Silvia Ghilezan.
\newblock The duality of classical intersection and union types.
\newblock {\em Fundamenta Informaticae}, 170(1-3):39--92, 2019.

\bibitem{dunfield2021bidirectional}
Jana Dunfield and Neel Krishnaswami.
\newblock Bidirectional typing.
\newblock {\em ACM Computing Surveys (CSUR)}, 54(5):1--38, 2021.

\bibitem{dunfield_pfenning_tridirectional}
Jana Dunfield and Frank Pfenning.
\newblock Tridirectional typechecking.
\newblock In {\em Proceedings of {POPL}}, 2004.

\bibitem{eades2017cointuitionistic}
Harley Eades~III and Gianluigi Bellin.
\newblock A cointuitionistic adjoint logic.
\newblock arXiv:1708.05896, 2017.

\bibitem{erdweg2015co}
Sebastian Erdweg, Oliver Bra{\v{c}}evac, Edlira Kuci, Matthias Krebs, and Mira
  Mezini.
\newblock A co-contextual formulation of type rules and its application to
  incremental type checking.
\newblock In {\em Proceedings of the 2015 ACM SIGPLAN International Conference
  on Object-Oriented Programming, Systems, Languages, and Applications}, pages
  880--897, 2015.

\bibitem{felleisen_etal_calculus_assignments}
Mattias Felleisen and Daniel Friedman.
\newblock A calculus for assignments in higher-order languages.
\newblock In {\em Proceedings of {POPL}}, 1987.

\bibitem{spj-scrap-applications}
Barry Jay and Simon~Peyton Jones.
\newblock Scrap your type annotations.
\newblock In {\em Proceedings of {Mathematics of Program Construction}}, 2008.

\bibitem{semanticdomain}
Neel Krishnaswami.
\newblock Polarity and bidirectional typechecking.
\newblock Available at
  \url{https://semantic-domain.blogspot.com/2018/08/polarity-and-bidirectional-typechecking.html},
  2018.

\bibitem{krishnaswami-inverting}
Neel Krishnaswami.
\newblock Inverting bidirectional typechecking.
\newblock Available at
  \url{https://semantic-domain.blogspot.com/2019/05/inverting-bidirectional-typechecking.html},
  2019.

\bibitem{conor}
Conor McBride.
\newblock Basics of bidirectionalism.
\newblock Available at
  \"https://pigworker.wordpress.com/2018/08/06/basics-of-bidirectionalism/".

\bibitem{mercer2022implicit}
Henry Mercer, Cameron Ramsay, and Neel Krishnaswami.
\newblock Implicit polarized {F}: local type inference for impredicativity.
\newblock arXiv:2203.01835, 2022.

\bibitem{miquey2019classical}
{\'E}tienne Miquey.
\newblock A classical sequent calculus with dependent types.
\newblock {\em ACM Transactions on Programming Languages and Systems (TOPLAS)},
  41(2):1--47, 2019.

\bibitem{munch2009focalisation}
Guillaume Munch-Maccagnoni.
\newblock Focalisation and classical realisability.
\newblock In {\em International Workshop on Computer Science Logic}, pages
  409--423. Springer, 2009.

\bibitem{munch2013syntax}
Guillaume Munch-Maccagnoni.
\newblock {\em Syntax and Models of a non-Associative Composition of Programs
  and Proofs}.
\newblock PhD thesis, Universit{\'e} Paris-Diderot-Paris VII, 2013.

\bibitem{ostermann2022introduction}
Klaus Ostermann, David Binder, Ingo Skupin, Tim S{\"u}berkr{\"u}b, and Paul
  Downen.
\newblock Introduction and elimination, left and right.
\newblock {\em Proceedings of the ACM on Programming Languages},
  6(ICFP):438--465, 2022.

\bibitem{pruiksma2018adjoint}
Klaas Pruiksma, William Chargin, Frank Pfenning, and Jason Reed.
\newblock Adjoint logic.
\newblock {\em Unpublished manuscript, April}, 2018.

\bibitem{spiwack2014dissection}
Arnaud Spiwack.
\newblock A dissection of {L}.
\newblock {\em Unpublished draft (cit. on pp. 121, 129)}, 2014.

\bibitem{tsukada2018intersection}
Takeshi Tsukada and Koji Nakazawa.
\newblock Intersection and union type assignment and polarised
  $\lambda$$\mu$$\mu$.
\newblock {\em Draft, University of Tokyo}, 2018.

\bibitem{zeilberger2015balanced}
Noam Zeilberger.
\newblock Balanced polymorphism and linear lambda calculus.
\newblock {\em Talk at TYPES}, 2015.

\end{thebibliography}

\end{document}